\documentclass{iopconfser}
\usepackage{amsmath}  
\usepackage{amsfonts} 

\usepackage{a4wide,amssymb,cite,graphicx}
\usepackage{epsfig}
\usepackage{hyperref}

\begin{document}

\noindent Conference Proceedings for BCVSPIN 2024: Particle Physics and Cosmology in the Himalayas\\Kathmandu, Nepal, December 9-13, 2024 

\title{Precision measurements of weak interaction parameters at Belle and Belle II}

\author{Ansu Johnson$^{1}$, on behalf of the Belle II Collaboration}

\affil{$^1$Department of Physics, Indian Institute of Technology, Madras, India}

\email{ansujohnson@physics.iitm.ac.in}

\begin{abstract}
We report on $CP$ violation measurements at Belle and Belle II, focusing on improving the precision of observables sensitive to the Cabbibo-Kobayashi-Maskawa matrix parameters. We present recent measurements of the CKM unitarity angle $\phi_2 (\alpha)$ using the $B \to \pi^0 \pi^0$ and $B^0 \to \rho^+ \rho^-$ decay channels, along with the first combined $\phi_3 (\gamma)$ measurement from Belle and Belle II. Additionally, we discuss the latest determinations of the CKM matrix elements $|V_{cb}|$ and $|V_{ub}|$ through exclusive decay modes $B \to D^{(*)} \ell \nu_\ell$ and $B \to \pi \ell \nu_\ell$, respectively. These results demonstrate improved precision and remain consistent with Standard Model predictions.

\end{abstract}

\section{Introduction}
Charge-parity (\(CP\)) violation in the quark sector is a fundamental aspect of the Standard Model (SM). It plays a crucial role in explaining the matter-antimatter asymmetry observed in the universe. Within the SM, \(CP\) violation arises from a single irreducible complex phase in the Cabibbo-Kobayashi-Maskawa (CKM) matrix~\cite{Cabibbo:1963yz,Kobayashi:1973fv}, which dictates quark mixing and weak decays. Precise measurements of \(CP\) violation and CKM parameters provide essential tests of the SM and impose constraints on potential physics beyond SM via loop-induced processes.  

The Belle and Belle II~\cite{BelleII:2010} experiments, conducted at the asymmetric-energy electron-positron collider SuperKEKB~\cite{SuperKEKB:2018} in Tsukuba, Japan, are designed to explore heavy flavor physics with unprecedented precision. The Belle experiment, which collected approximately 1~ab\(^{-1}\) of data at the \(\Upsilon(4S)\) resonance, played a pivotal role in confirming the CKM mechanism as the primary source of \(CP\) violation within the SM. Its successor, Belle II, aims to surpass Belle’s sensitivity by accumulating a dataset of 50~ab\(^{-1}\), leveraging improved detector performance and significantly higher luminosity. Since the commencement of data collection in March 2019, SuperKEKB has reached a peak luminosity of \(5.1 \times 10^{34}\)~cm\(^{-2}\)s\(^{-1}\), more than twice the maximum achieved by Belle. As of now, Belle II has recorded a total data sample of 575~fb$^{-1}$. The results presented here are based on 365~fb$^{-1}$ of data collected during Belle II Run-1 at the $\Upsilon(4S)$ resonance.



\section{Combined Belle and Belle II measurement of angle $\phi_{3}$ ($\gamma$)}
The angle $\phi_3 = \arg \left( -V_{ud} V_{ub}^* / V_{cd} V_{cb}^* \right)$ can be determined through the interference of $b \to c$ and $b \to u$ processes that lead to the same final state. These processes are predominantly tree-level with no significant contributions from physics beyond the Standard Model. We present the first combination of all Belle and Belle II measurements, incorporating different methods and utilizing data samples of varying sizes, as detailed in Table~\ref{table:phi_3}. The sensitivity is mainly driven by the BPGGSZ method~\cite{Abudinen2022_BPGGSZ,Resmi2019_BPGGSZ,Poluektov2010_BPGGSZ}. The final combined result~\cite{phi3:combo} gives $\phi_3 = (75.2 \pm 7.6)^\circ$, which is in agreement with the current world average, $\phi_3^{\text{WA}} = (66.4^{+2.8}_{-3.0})^\circ$~\cite{PDG:2022}.

\begin{table}[h!]   
\centering
\resizebox{\textwidth}{!}{
\begin{tabular}{c|c|c|c|c}
\hline \hline$B$ decay & $D$ decay & Method & \begin{tabular}{c} 
Data set [fb $\left.{ }^{-1}\right]$ \\
$($ Belle + Belle II $)$
\end{tabular} & Ref. \\
\hline$B^{+} \rightarrow D h^{+}$ & $D \rightarrow K_S^0 \pi^0, K^{-} K^{+}$ & GLW & $711+189$ & \cite{Adachi2024_GLW} \\
\hline$B^{+} \rightarrow D h^{+}$ & $D \rightarrow K^{+} \pi^{-}, K^{+} \pi^{-} \pi^0$ & ADS & $711+0$ & \cite{Horii2011_ADS,Nayak2013_ADS} \\
\hline$B^{+} \rightarrow D h^{+}$ & $D \rightarrow K_S^0 K^{-} \pi^{+}$ & GLS & $711+362$ & \cite{Adachi2023_GLS} \\
\hline$B^{+} \rightarrow D h^{+}$ & $D \rightarrow K_S^0 h^{-} h^{+}$ & BPGGSZ & $711+128$ & \cite{Abudinen2022_BPGGSZ} \\
\hline$B^{+} \rightarrow D h^{+}$ & $D \rightarrow K_S^0 \pi^{-} \pi^{+} \pi^0$ & BPGGSZ & $711+0$ & \cite{Resmi2019_BPGGSZ} \\
\hline$B^{+} \rightarrow D^* K^{+}$ & $D^{*} \rightarrow D\pi^0$  & GLW
& $210+0$ & \cite{Abe2006_GLW} \\
 & $D \rightarrow K_S^0 \pi^0, K_S^0 \phi, K_S^0 \omega, K^{-} K^{+}, \pi^{-} \pi^{+}$ & & & \\
 \hline$B^{+} \rightarrow D^* K^{+}$ & $D^{*} \rightarrow D\pi^0,D\gamma$  & BPGGSZ
& $605+0$ & \cite{Poluektov2010_BPGGSZ} \\
 & $D \rightarrow K_S^0 \pi^{+}\pi^{-}$ & & & \\
\hline \hline
\end{tabular}}
\caption{Belle and Belle II measurements used for the combined measurement of $\phi_3$.}
\label{table:phi_3}
\end{table}


\section{Determination of $\phi_{2}$ ($\alpha$)}

The angle $\phi_2 = \arg\left(-{V_{ud}V_{ub}^*}/{V_{td}V_{tb}^*}\right)$ is extracted from time-dependent $CP$-violation parameters in $b \to u \bar{u} d$ decays. In $e^+e^-$ collisions at the $\Upsilon(4S)$ resonance, pairs of $B^0$ and $\bar{B}^0$ mesons are produced through the decay $\Upsilon(4S) \to B^0 \bar{B}^0$. When the signal $B$ meson ($B_{\text{\rm sig}}^0$) decays into a flavor eigenstate and the tagging meson ($B_{\text{\rm tag}}^0$) decays with flavor $q$ ($q = +1$ for $B^0$ and $q = -1$ for $\bar{B}^0$), $CP$-violation manifests as a time difference, $\Delta t$, between the two $B$ mesons. The flavor of the $B_{\text{\rm tag}}^0$ is determined by a graph-neural-network flavor tagger (GFLAT)~\cite{GNN:2024}, which enhances the effective tagging efficiency to $(37.40 \pm 0.43 \pm 0.36)\%$, compared to the $(31.68 \pm 0.45)\%$ efficiency achieved by a category-based flavor tagger.



In the absence of loop contributions from $b \to u \bar{u} d$ decays, such as $B^0 \to \rho^+ \rho^-$ or $B^0 \to \pi^+ \pi^-$, $\phi_2$ can be determined via the relation $S = 2 \sin \phi_2$. However, if a non-negligible $b \to d$ loop amplitude exists, the parameters $S$ and $C$ are modified to $S = \sin(2 \phi_2 + 2 \Delta \phi_2)$ and $C \neq 0$. Isospin analysis can be used to account for the loop amplitude. In the case of $B \to \pi \pi$ decays, this analysis involves the decays $B \to \pi^+ \pi^-, \pi^0 \pi^0$, and $\pi^+ \pi^0$, similarly it can be applied to $B \to \rho \rho$. Comparing $B \to \rho \rho$ with $B \to \pi \pi$, the loop amplitude for $B^0 \to \rho \rho$ is significantly smaller, offering improved sensitivity to $\phi_2$.

\subsection{Measurement of the Branching Fraction and $A_{CP}$ of $B^0 \to \pi^0 \pi^0$}
\begin{figure}
    \centering
    \includegraphics[height=4.5cm]{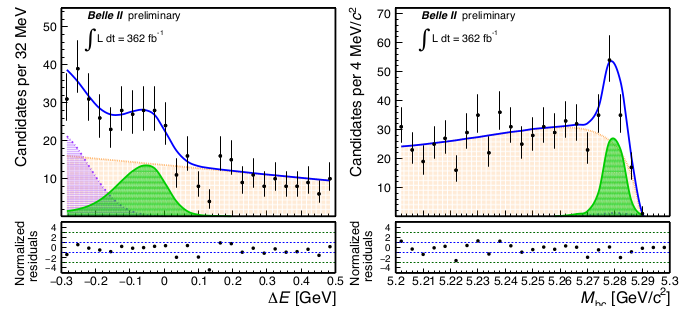}
    \caption{$\Delta E$ and $M_{\rm bc}$ distribution for $B^0 \to \pi^0 \pi^0$ decays. The black dots represent the data, the blue curve represents the total fit result, the green area represents the signal, the purple area represents the $B \bar{B}$ backgrounds, and the orange area corresponds to the $q \bar{q}$ backgrounds.}
    \label{fig:alpha_pi0pi0}
\end{figure}

We measure the branching fraction and direct $CP$ asymmetry in $B^0 \to \pi^0 \pi^0$ decays\cite{pi0pi0}. The $\pi^0 \to \gamma \gamma$ candidates suffer from significant backgrounds due to hadronic clusters and beam-induced processes. To suppress mis-reconstructed photons, a dedicated machine-learning algorithm using a Boosted Decision Tree (BDT) is employed. The signal is extracted by fitting the distributions of $\Delta E$, $M_{\text{\rm bc}}$, the transformed $q\bar{q}$ suppression BDT output, and the transformed wrong-flavor tagging probability. We obtain a branching fraction of $B = 1.26 \pm 0.20 \pm 0.12$, where the uncertainties are statistical and systematic, and a direct $CP$ asymmetry of $C = -0.06 \pm 0.30 \pm 0.05$, consistent with the current world average. The fitted distributions of $\Delta E$ and $M_{\rm bc}$ are shown in Figure~\ref{fig:alpha_pi0pi0}.

\subsection{Measurement of $B^0 \to \rho^+ \rho^-$}
We present the measurement of the polarized $B^0 \to \rho^+ \rho^-$ decays~\cite{rhorho}. The longitudinally polarized state is $CP$-even, while the transversely polarized state contains both $CP$-even and $CP$-odd components. We perform an angular analysis to distinguish these states and measure the $CP$ violation parameter for the longitudinal polarization. The signal is extracted using the distributions of $\Delta E$, $m(\pi^\pm \pi^0)$, the cosine of the helicity angle for $\rho^\pm$, and the transformed $q \bar{q}$ suppression output. We measure a branching fraction $(29.0^{+2.3}_{-2.2} {}^{+3.1}_{-3.0}) \times 10^{-6}$ and a longitudinal polarization fraction $0.921^{+0.024}_{-0.025} {}^{+0.017}_{-0.015}$, both consistent with the current world average.

\begin{figure}[h]
    \centering
    \includegraphics[height=4.5cm]{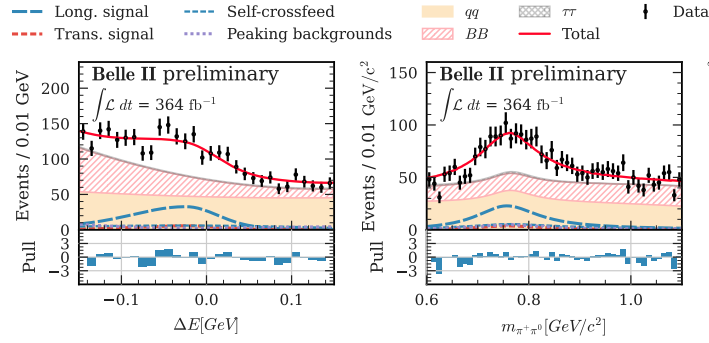}
    \hspace{2mm}
    \includegraphics[height=4.5cm]{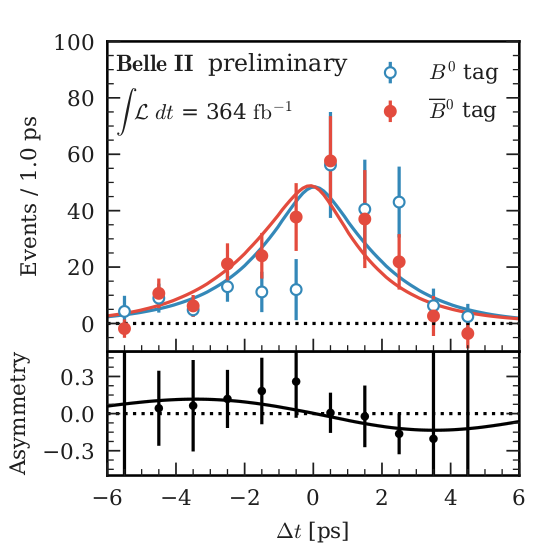}
    \caption{Distributions for $\Delta E$ (left) and $m_{\pi^{+}\pi^{0}}$ (center) in $B^{0} \to \rho^{+}\rho^{-}$ decays. The background-subtracted $\Delta t$ distributions for $B^{0}$ tag and $\bar B^{0}$ tag, as well as their asymmetry (lower panel) are shown on the right plot.}
    \label{fig:alpha_rhorho}
\end{figure}

We then measure the time-dependent $CP$ asymmetry from the $\Delta t$ distribution, where the flavor of $B^0_{\text{tag}}$ is identified using GFlaT. The CP-violating parameters are determined as $S = -0.26 \pm 0.19 \pm 0.08$ and $C = -0.02 \pm 0.12 {}^{+0.06}_{-0.05}$, which are consistent with the world average. The optimized selection criteria and the use of GFlaT enhance the sensitivity of these measurements. The distributions for $\Delta E$, $m_{\pi^{+}\pi^{0}}$ and $\Delta t$ are shown in Figure~\ref{fig:alpha_rhorho}.

We extract the value of $\phi_2$ using the updated $B^0 \to \rho^+ \rho^-$ result. An isospin analysis, incorporating the current world averages for $B \to \rho \rho$ decays, gives $\phi_2 = (91.5^{+4.5}_{-5.4})^\circ$. This result is slightly less precise than that obtained in \cite{Charles:2017} due to corrections to the world-average branching fractions of $B \to \rho \rho$, which are based on the $B^+$ and $B^0$ production ratio at the $\Upsilon(4S)$ resonance, as measured in \cite{Belle:2023}. Combining the world average with our $B^0 \to \rho^+ \rho^-$ results, we obtain a more precise value of $\phi_2 = (92.6^{+4.5}_{-4.8})^\circ$, improving the precision by 6\%. The dominant uncertainty in $\phi_2$ comes from the time-dependent $CP$ asymmetry parameter $S$ in the $B^0 \to \rho^+ \rho^-$ and $B^0 \to \rho^0 \rho^0$ decays.

\section{Determination of $|V_{cb}|$ using $B^- \to D^* \ell \bar{\nu}$ }
With the full Belle dataset, a comprehensive measurement of the complete set of angular coefficients 
for the exclusive $B^- \to D^* \ell \bar{\nu}$ decay is performed~\cite{Prim:2023angular}. One $B$ meson is reconstructed using hadronic decays, while the other $B$ is reconstructed in the decay channels 
$B^0 \to D^{*+} \ell \bar{\nu}$ with $D^{*+} \to D^0 \pi^+$ / $D^+ \pi^0$, and  
$B^- \to D^{*0} \ell \bar{\nu}$ with $D^{*0} \to D^0 \pi^0$. 
Non-resonant $e^+ e^-$ interactions are suppressed using a multivariate classifier.  

The angular coefficients are extracted from data, binned according to the hadronic recoil parameter  
$w = {(m_B^2 + m_{D^*}^2 - q^2)}/{2 m_B m_{D^*}}$, separately for $\ell = e, \mu$ and $B = B^0, B^+$ modes. 
Within each $w$ bin, the signal yields are determined in bins of the decay angles $\theta_\ell$, $\theta_V$, and $\chi$. 
Here, $\theta_\ell$ represents the angle between the lepton and the direction opposite to the $B$ meson in the virtual $W$-boson rest frame, $\theta_V$ is the angle between the $D$ meson and the direction opposite to the $B$ meson in the $D^*$ rest frame, and $\chi$ denotes the angle between the decay planes spanned by the $W^- \ell$ and $D^*-D$ systems in the $B$ meson rest frame.  

The extracted angular coefficients facilitate the determination of the form factors governing the $B \to D^*$ transition and the magnitude of $|V_{cb}|$. Using recent lattice QCD calculations for the form factors, the value of $|V_{cb}|$ is found to be $(41.0 \pm 0.7) \times 10^{-3}$ based on the Boyd-Grinstein-Lebed (BGL) parameterization~\cite{Boyd:1997}.  
This result is consistent with the fit to the one-dimensional differential spectra derived from the same dataset~\cite{Prim:2023differential}, as well as with the most precise determinations obtained from inclusive $B \to X_c \ell \nu$ decays~\cite{Bordone:2021,Bernlochner:2022,Finauri:2024}.  
Furthermore, possible lepton flavor universality violation is examined by comparing the angular distributions of electrons and muons, with no significant deviation from Standard Model expectations observed.

\section{$|V_{ub}|$ from simultaneous measurements of untagged $B^0 \to \pi^- \ell^+ \nu$ and $B^+ \to \rho^0 \ell^+ \nu$ decays}

This analysis uses a data sample of 387 million $B\bar{B}$ meson pairs recorded by Belle II~\cite{Adachi:2024}. The decays $B^0 \to \pi^- \ell^+ \nu$ and $B^+ \to \rho^0 \ell^+ \nu$ are reconstructed without identifying the partner $B$ meson. Events are categorized into 13 intervals for the pion mode and 10 intervals for the rho mode based on the squared momentum transfer $q^2$. The signal yields for both modes are extracted simultaneously using a two-dimensional fit in $\Delta E$ and $M_{\text{bc}}$ for each $q^2$ bin. This method effectively accounts for cross-feed between the two decay modes.

The partial branching fractions are derived from the fitted signal yields after applying efficiency corrections as a function of $q^2$. The total branching fraction is computed by summing the partial branching fractions, accounting for systematic correlations. The preliminary results yield total branching fractions of $\mathcal{B}(B^0 \to \pi^- \ell^+ \nu_\ell) = (1.516 \pm 0.042  \pm 0.059)  \times 10^{-4}$ and $\mathcal{B}(B^+ \to \rho^0 \ell^+ \nu_\ell) = (1.625 \pm 0.079  \pm 0.180 ) \times 10^{-4}$. These values are consistent with world averages, and the precision is comparable to previous measurements from Belle and BaBar.

To extract $|V_{ub}|$, the decay form factors for $B^0 \to \pi^- \ell^+ \nu$ are parameterized using the Bourrely-Caprini-Lellouch (BCL) model~\cite{Bourrely:2009}, while the Bharucha-Straub-Zwicky (BSZ) parametrization~\cite{Bharucha:2016} is applied for $B^+ \to \rho^0 \ell^+ \nu$. By fitting the measured partial branching fractions of $B^0 \to \pi^- \ell^+ \nu$ as functions of $q^2$, and incorporating constraints on non-perturbative hadronic contributions from lattice QCD calculations~\cite{Aoki:2022}, we obtain the preliminary result $|V_{ub}| = (3.93 \pm 0.09 \pm 0.13 \pm 0.19) \times 10^{-3}$, where the third uncertainty is theoretical. Similarly, the preliminary result from the $B^+ \to \rho^0 \ell^+ \nu$ decay, including the constraints from the light-cone sum rule (LCSR)\cite{Bharucha:2016}, is found to be $|V_{ub}| = (3.19 \pm 0.12  \pm 0.17 \pm 0.26 ) \times 10^{-3}$.

\begin{figure}[h!]
    \centering
    \includegraphics[height=4.5cm]{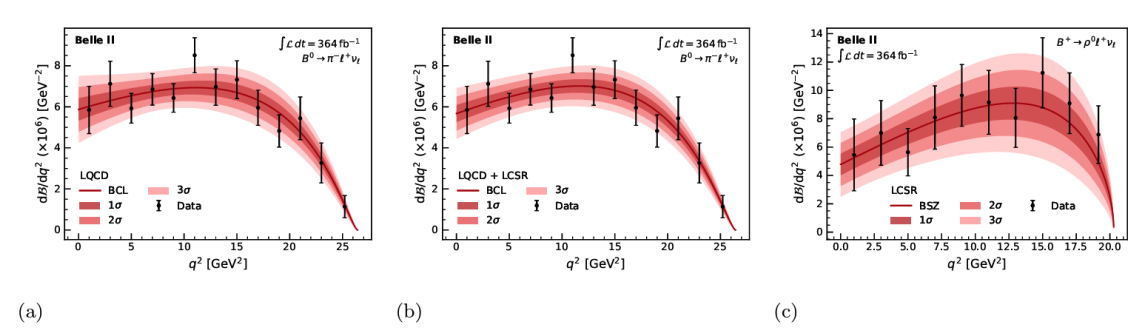}
    \caption{The measured partial branching fractions as a function of $q^2$ for $B^0 \to \pi^- \ell^+ \nu$ (a, b) and $B^+ \to \rho^0 \ell^+ \nu$ (c). The fitted differential rates are displayed alongside the uncertainty bands corresponding to one, two, and three standard deviations, with the form factors constrained by (a) LQCD, (b) LQCD and LCSR, and (c) LCSR predictions.}
    \label{fig:Vub}
\end{figure}

The $|V_{ub}|$ value obtained from the $B^0 \to \pi^- \ell^+ \nu$ mode is determined as $|V_{ub}| = (3.93 \pm 0.09  \pm 0.13  \pm 0.19 ) \times 10^{-3}$, which is consistent with previous exclusive measurements. The result from the $B^+ \to \rho^0 \ell^+ \nu$ mode is lower but remains compatible with previous experimental determinations from $B \to \rho\ell\nu$ decays. In both cases, the precision is limited by theoretical uncertainties. Figure~\ref{fig:Vub} shows the measured and fitted differential rates of $B^0 \to \pi^- \ell^+ \nu$ and $B^+ \to \rho^0 \ell^+ \nu$, as well as the one, two, and three standard deviations from the fits.

\section{Summary}
We report on recent measurements of the CKM Unitarity Triangle angles and sides using $B$ decays, including the first combined Belle and Belle II determination of the CKM angle $\phi_3$ and new measurements of $\phi_2$ from $B \to \pi^0 \pi^0$ and $B \to \rho^+ \rho^-$ decays. Additionally, we discuss the extraction of $|V_{cb}|$ and $|V_{ub}|$ from exclusive decays. All results align with previous measurements, enhancing the precision of CKM parameter determinations.

\bibliographystyle{jhep}
\bibliography{ref}{}

\end{document}